\documentclass[reprint,prd,aps,twocolumn,amsmath,amssymb,nofootinbib,superscriptaddress]{revtex4-2}
\usepackage{amsfonts}
\usepackage{bm}% bold math
\usepackage{mathtools}
\usepackage{stackrel}
\usepackage[utf8]{inputenc}
\usepackage{blindtext}
\usepackage{float}
\restylefloat{table}
\usepackage{booktabs}
\usepackage{enumitem}
\usepackage{epsfig}
\usepackage[dvipsnames]{xcolor}
\usepackage{etoolbox} % for \appto
\usepackage{lipsum} % for mock text
\usepackage{verbatim}
\usepackage[colorlinks,linkcolor=blue]{hyperref}

\hypersetup{pdfstartview=FitV,colorlinks=true,linkcolor=blue,citecolor=magenta,filecolor=midblue,urlcolor=blue}

\definecolor{midblue}{rgb}{0,0,0.5}
\usepackage{dcolumn}% Align table columns on decimal point
\usepackage{bm}% bold math
\usepackage{subcaption}

\def\Ga{\Gamma}

\def\pt#1{\phantom{#1}}

\def\3g#1#2#3{^{(3)}\Ga^{#1}_{\pt{#1}#2#3}}

\newcommand{\beq}{\begin{equation}}
\newcommand{\eeq}{\end{equation}}
\newcommand{\bea}{\begin{eqnarray}}
\newcommand{\eea}{\end{eqnarray}}

\begin{document}

\preprint{number}

\title{Primordial observables of explicit diffeomorphism violation in gravity}% Force line breaks with \\

\author{Mohsen Khodadi}\email{khodadi@kntu.ac.ir}
\affiliation{School of Physics, Institute for Research in Fundamental Sciences (IPM),\\ P.~O.~Box 19395-5531, Tehran, Iran}
\affiliation{School of Physics, Damghan University, Damghan 3671641167, Iran}
\affiliation{Center for Theoretical Physics, Khazar University, 41 Mehseti Str., AZ1096 Baku, Azerbaijan}

\author{Nils A. Nilsson}
 \email{nilsson@ibs.re.kr}
\affiliation{%
Cosmology, Gravity, and Astroparticle Physics Group, Center for Theoretical Physics of the Universe, Institute for Basic Science (IBS), Daejeon, 34126, Korea
}%
\affiliation{%
LTE, Observatoire de Paris, Université PSL, CNRS, LNE, Sorbonne Universit\'e, 61 avenue de l’Observatoire, 75014 Paris, France}%

\author{Gaetano Lambiase}
\email{lambiase@sa.infn.it}
\affiliation{Dipartimento di Fisica ‘‘E.R. Caianiello’’, Università di Salerno, Via Giovanni Paolo II, I-84084 Fisciano (SA), Italy}

\affiliation{INFN, Gruppo collegato di Salerno, Via Giovanni Paolo II, I-84084 Fisciano (SA), Italy}

\author{Javad T. Firouzjaee}\email{firouzjaee@kntu.ac.ir}
\affiliation{Department of Physics, K. N. Toosi University of Technology,	P. O. Box 15875-4416, Tehran, Iran}

\date{\today}% It is always \today, today,

\begin{abstract}
We investigate the potential for current and future gravitational-wave detectors to observe imprints of explicit diffeomorphism violation in primordial signals. Starting from a simple model with known effects, we derive the strain amplitude and power spectrum for primordial gravitational waves, both of which are affected by the symmetry breaking. Through this, we directly find predictions for the tensor spectral index and tensor-to-scalar which are different from general relativity. By considering the known sensitivity curves for NANOGrav, SKA, THEIA, $\mu$-ARES, ASTROD-GW, LISA, BBO, DECIGO, CE, AION-km, AEDGE, ET, and aLIGO, we place observability limits on the parameters controlling the diffeomorphism violation. For instance, we find that aLIGO could observe signals for \(s_{00} \lesssim -0.1\), while more sensitive future detectors like LISA and DECIGO could probe violations as small as \(s_{00} \approx -5 \times 10^{-4}\) and \(-3 \times 10^{-3}\), respectively. Finally, we consider the existing constraints on the number of relativistic degrees of freedom $\Delta N_{\rm eff}$ which is tightly constrained by Big-Bang Nucleosynthesis (BBN) and we find that $\Delta N_{\rm eff}$ only weakly depends on the symmetry breaking but places a lower bound on the coefficients which is consistent with available bounds from the speed of gravitational waves.
\end{abstract}

\maketitle

%\tableofcontents

\section{\label{sec:intro}Introduction}
It is generally expected that general relativity and the standard model do not constitute the fundamental description of Nature, but that they are low-energy limits of a single theory which are accurate at the scales available for measurement. This is motivated by the expectation that there exists a single, all-encompassing theory which describe the known fundamental interactions. This implies, for example, the existence of a renormalizable theory of gravity from which general relativity (GR) emerges as a low-energy limit, with the ultraviolet cutoff scale usually being assumed to be the Planck scale. Any theory attempting to connect general relativity and the standard model should in principle be accurate up to this scale, and searching for deviations from theoretical predictions in high-energy observables (or as remnants in infrared physics) is a promising approach when searching for new physics. One possible such signal is the deviation from the fundamental spacetime symmetries such as local Lorentz and diffeomorphism symmetry, and the possibility of {\it spacetime-symmetry breaking}, as has been shown to emerge from for example string field theory \cite{Kostelecky:1988zi,Kostelecky:1989jw}. In general, the symmetry breaking can manifest in two distinct ways: spontaneously or explicitly, and is signaled by the presence of one or more non-dynamical background tensor fields, and the origin of the tensor fields (explicit or spontaneous breaking) can significantly affect the phenomenology. The difference between these can be subtle in practice and has been discussed at length elsewhere~\cite{Bluhm:2004ep,Bluhm:2014oua,Bluhm:2021lzf}. The discovery of consistent solutions in string field theory instigated a large research effort which to date has placed a large number of strong constraints across all sectors of the standard model and GR, using data from a vast number of experiments and observations in the past decades~\cite{Kostelecky:2008ts}\footnote{Updated annually.}. For instance, linearized-gravity tests have been performed in the Solar system~\cite{Iorio:2012gr, LePoncin-Lafitte:2016ocy} including Lunar-Laser Ranging (LLR)~\cite{Bourgoin:2017fpo,Bourgoin:2020ckq}, with pulsars~\cite{Shao:2014oha,Shao:2018vul}, and in short-range gravity~\cite{Bailey:2022wuv}. Moreover, the recent detection of gravitational waves (GWs) by the LIGO-Virgo-Kagra (LVK) collaboration has opened a new window to the Universe \footnote{The Hubble tension, which represents a discrepancy in the cosmic expansion rate between early and late cosmological epochs, is prone to opening a new window into physics beyond the standard model, including potential signatures of spacetime symmetry breaking in the early Universe \cite{Khodadi:2023ezj}.}, allowing us to probe gravity in an entirely new way; this has given rise to more limits on spacetime symmetries, see for example~\cite{Haegel:2022ymk,Wang:2021ctl,Nilsson:2022mzq,Ray:2023sbr,Liu:2020slm,LIGOScientific:2017zic,Mewes:2019dhj}. 

However, GW observations are not only relevant for late-time signals; indeed, observations can in principle be done of tensor signals emitted during inflation (or even prior to inflation in Big Bounce models~\cite{Malkiewicz:2020fvy,Li:2024dce}) or through phase transitions around Big-Bang Nucleosynthesis (BBN) time~\cite{Kosowsky:1991ua,Kamionkowski:1993fg} (see also \cite{Khodadi:2021ees,Capozziello:2018qjs,Khodadi:2018scn}), from topological defects~\cite{Vachaspati:1984gt,Figueroa:2012kw}, generation or merger of a population of primordial black holes~\cite{Garcia-Bellido:1996mdl,Dong:2015yjs,LISACosmologyWorkingGroup:2023njw} can all emit so-called primordial gravitational waves (PGWs), which is exactly the topic we consider in the present paper. Apart from the emission mechanisms mentioned above, a primary source of PGWs is cosmic inflation, the momentary period where the universe underwent exponential expansion~\cite{Maggiore:1999vm,Wang:2024gko} and which today is the standard solution to the horizon and flatness problems, as well as the small temperature anisotropies in the Cosmic Microwave Background (CMB)~\cite{Guth:1980zm,Kamionkowski:1996ks}. The inflationary mechanism allowed the primordial universe to grow by a factor of approximately $10^{60}$ on a timescale of around $10^{-36}$s \cite{Achucarro:2022qrl}, which allowed quantum fluctuations to expand past the Hubble horizon, upon which they froze and then re-entered the horizon as classical tensor perturbations (PWGs) at later times. Many inflationary models predict a nearly scale-invariant spectrum of PGWs (as for the temperature fluctuations) with a slightly red-tilted spectrum~\cite{Guzzetti:2016mkm,Vagnozzi:2023lwo,BICEP2:2018kqh,Tristram:2021tvh}, a constraint which was recently made tighter by the data release from the Atacama Cosmology Telescope (ACT)~\cite{ACT:2025tim}. PGWs that enter the horizon at the relevant scales leave imprints in the CMB which are directly measurable, primarily B-mode polarizations and anisotropies, making the CMB one of the primary targets for detecting signals of PGWs. However, direct detection with current and next generation gravitational-wave observatories is also possible, see for example~\cite{Chongchitnan:2006pe,Fumagalli:2021dtd}, and given its primordial origin, it is an excellent probe of new physics, which we exploit in this paper.

In this paper, we consider {\it explicit} spacetime-symmetry breaking where diffeomorphisms are broken at the level of the action by positing the existence of a non-dynamical tensor field which transforms trivially under particle diffeomorphisms~\cite{Kostelecky:2020hbb,Kostelecky:2003fs} but still respects local particle Lorentz symmetry. Such explicit symmetry breaking is distinct from {\it spontaneous} symmetry breaking in which the field takes some non-zero vacuum expectation value and produces Nambu-Goldstone modes and massive modes~\cite{Bluhm:2014oua}\footnote{Models based on spontaneous Lorentz symmetry breaking give rise to a rich phenomenology, for instance see Refs. \cite{Khodadi:2023yiw,Khodadi:2022dff,Khodadi:2021owg,Khodadi:2022pqh,Khodadi:2022mzt}.} (and was studied in the present context in e.g. \cite{Khodadi:2025wuw}); instead, a non-dynamical tensor field is placed directly into the action and has no associated dynamics. The phenomenology of such systems has been considered previously, for example in \cite{ONeal-Ault:2020ebv,Reyes:2021cpx} where the authors studied the Hamiltonian formulation of such theories. A number of no-go issues which arise in explicit-breaking scenarios were considered in~\cite{Bailey:2024zgr}, where it was found that certain no-go constraints can be evaded in the linearized limit. Furthermore, cosmological solutions were developed in \cite{Reyes:2022dil,Nilsson:2023exc}, as well as \cite{Nilsson:2022mzq} which considered the effect on primordial fields. In this paper, we follow a similar approach but focus on observability of such signals with current and next-generation gravitational-wave detectors; specifically, we find the observability limits forNANOGrav \cite{NANOGrav:2023gor}, the Square Kilometer Array (SKA) \cite{Janssen:2014dka}, THEIA \cite{Theia:2019non}, $\mu$-Ares \cite{Sesana:2019vho}, ASTRO-GW \cite{Ni:2012eh}, the Atom Interferometer Observatory and Network (AION-Km) \cite{Badurina:2019hst}, the LISA interferometer~\cite{LISA:2017pwj}, the DECi-hertz Interferometer Gravitational wave Observatory (DECIGO) \cite{Kawamura:2020pcg}, the Atomic Experiment for Dark Matter and Gravity Exploration (AEDGE) \cite{AEDGE:2019nxb}, the Big Bang Observer (BBO) \cite{Crowder:2005nr}, the Einstein Telescope (ET) \cite{Sathyaprakash:2012jk}, the Cosmic Explorer (CE) \cite{Evans:2021gyd}, and Advanced LIGO (aLIGO) \cite{LIGOScientific:2014pky}. Finally, we consider bounds from BBN.

This paper is organized as follows: in Section~\ref{sec:theory} we introduce the cosmological model of following from explicit symmetry breaking; in Section~\ref{sec:perts}, we compute the power spectrum and observables from the PGWs; in Section~\ref{sec4} we compute the observability of the signals in current and future GW detectors; in Section~\ref{sec:BBN} we complement the gravitational-wave constraints by considering BBN bounds on the number of effective relativistic degrees of freedom, and we discuss and conclude in Section~\ref{sec:disc}.

Throughout this paper, we use units in which $c=1=\hbar$, and the metric signature $(-,+,+,+)$. We use Greek letters $\mu,\nu,\alpha,\hdots$ as spacetime indices and mid-alphabet Latin letters $i,j,k,\hdots$ as spatial indices.

\section{Cosmological background solution\label{sec:theory}}
At mass-dimension \( d \leq 4 \), the Lagrange density takes the form  
\begin{equation}\label{eq:lagr}
\mathcal{L} = \frac{\sqrt{-g}}{2\kappa}\left[ R + (k_{\rm R})^{\alpha\beta\mu\nu} R_{\alpha\beta\mu\nu}\right] + \mathcal{L}^\prime,
\end{equation}  
where $\kappa=8\pi G$, \( R_{\alpha\beta\mu\nu} \) is the Riemann tensor, and \( (k_{\rm R})^{\alpha\beta\mu\nu} \) are symmetry-breaking coefficients. These decompose into three distinct components: scalar term \(uR \), a trace-free term \( s^{\mu\nu} R_{\mu\nu}^{(\rm T)} \) (with \( R_{\mu\nu}^{(\rm T)} \) being the trace-free Ricci tensor), and \( t^{\alpha\beta\mu\nu} W_{\alpha\beta\mu\nu} \) (where \( W_{\alpha\beta\mu\nu} \) denotes the Weyl tensor). The term $\mathcal{L}^\prime$ contains kinetic and potential terms for $k_R$ in the case of spontaneous breaking, but it vanishes in the present context.

The terms \( u \), \( s^{\mu\nu} \), and \( t^{\alpha\beta\mu\nu} \) explicitly break particle diffeomorphism symmetry (but preserves local Lorentz symmetry) and observer covariance, and are non-dynamical background fields with predetermined geometric properties. Notably, the scalar term \( u \) can be removed by redefining the gravitational constant to generate Brans-Dicke theory and is generally mapped to scalar-tensor theory~\cite{Reyes:2022mvm, Bailey:2024zgr}; alternatively, in the case of constant $u$, such a redefinition renders it unobservable. This situation contrasts sharply with spontaneous symmetry breaking, where symmetry-breaking coefficients arise as vacuum expectation values of dynamical fields, fully governed by their own equations of motion. In the spontaneous scenario, the action remains invariant under particle transformations, and the theory naturally incorporates Nambu-Goldstone modes alongside massive modes, reflecting its dynamical structure \cite{Bluhm:2021lzf,Bluhm:2019ato}.  

Since the \( u \)-term can always be absorbed into the trace component of \( s^{\mu\nu}_T \), where it denotes the trace-free part of $s_{\mu\nu}$, we redefine the coefficient as  
\begin{equation}
k_{\rm R} = s_{\mu\nu} R^{\mu\nu} + t^{\alpha\beta\mu\nu} W_{\alpha\beta\mu\nu},
\end{equation}  
where \( s_{\mu\nu} \) now includes a trace contribution and is no longer trace-free. This redefinition will be adopted consistently in the subsequent analysis. Under particle and observer diffeomorphisms along $\xi^\mu$, the metric $g_{\mu\nu}$ and $s_{\mu\nu}$ ($t^{\alpha\beta\mu\nu}$ follows similar transformation rules as $s_{\mu\nu}$) transform as~\cite{Bluhm:2014oua,Kostelecky:2020hbb}
\begin{equation}
    \begin{aligned}
    &\text{Particle: } &g_{\mu\nu} \to g_{\mu\nu} - \mathcal{L}_\xi g, \quad &s_{\mu\nu} \to s_{\mu\nu},\\
    &\text{Observer: } &g_{\mu\nu} \to g_{\mu\nu} + \mathcal{L}_\xi g, \quad &s_{\mu\nu} \to s_{\mu\nu}+\mathcal{L}_\xi s,
    \end{aligned}
\end{equation}
and we therefore identify observer diffeomorphisms with general coordinate transformations, under which the Lagrangian (\ref{eq:lagr}) is invariant. Due to the explicit nature of the symmetry breaking, local Lorentz symmetry is still preserved~\cite{Kostelecky:2020hbb}. Note also that the placement of indices is important in the explicit breaking context: considering for example that $s_{\mu\nu}$ with covariant indices is fixed under particle diffeomorphisms, the quantity $s^\alpha_{~\nu}=g^{\mu\alpha}s_{\mu\nu}$ is no longer fixed, since the metric is a dynamical field and transforms non-trivially; in what follows, we consider $s_{\mu\nu}$ with covariant indices to be fixed. Other index configurations can of course be considered and are tantamount to considering different theories. Finally, we introduce the simplification $t_{\alpha\beta\mu\nu}=0$ and do not consider it in the analysis below, since we are only interested in coefficients which affect the expansion history as well as perturbations, and since the Weyl tensor vanishes around an FLRW background, this term only becomes important at the level of perturbations; as such, we do not consider it further, and instead leave it for future, more general work.

When spacetime symmetries are explicitly broken, serious complication with the Bianchi identities normally arise, which is known as the no-go result~\cite{Kostelecky:2003fs,Bailey:2024zgr}. In the case of $s_{\mu\nu}$, the constraint equation arising from the Bianchi identities reads~\cite{ONeal-Ault:2020ebv,Nilsson:2022mzq}
\begin{equation}\label{eq:bianchi}
    \nabla_\mu(T_s)^\mu_{~\nu}=\frac{1}{2}R^{\mu\lambda}\nabla_\nu s_{\mu\lambda}-\nabla_\mu(R^{\mu\lambda}s_{\lambda\nu}),
\end{equation}
where $T_s$ consists of all terms in the stress-energy tensor proportional to $s_{\mu\nu}$. In order to be compatible with Riemannian geometry, any ansatz for $s_{\mu\nu}$ must satisfy the constraints arising from the above relation: $(T_s)^\mu_{~\nu}$ can be separately conserved or considered as part of the matter action, at which point the whole right-hand side of the Einstein equations must be conserved as a whole. Let us now introduce the Friedmann-Lemaître-Robertson-Walker (FLRW) metric as
\begin{equation}\label{eq:flrw}
	ds^2 = -dt^2 + a^2(t)\left(\frac{dr^2}{1-kr^2} + r^2\,d\Omega_2^2\right),
\end{equation}
where \( a(t) \) is the cosmic scale factor, $d\Omega_2^2$ the line element on the 2-sphere, \( k \) denotes the spatial curvature parameter, taking values \( k = \{+1, 0, -1\} \) for closed, flat, and open universes, respectively. While we compute all background quantities using the general FLRW metric, we subsequently specialize to the spatially flat case (\( k = 0 \)). In these coordinates, the spatial part $\nabla_\mu(T_s)^\mu_{~j}$ turns out to be proportional to $\partial_js_{00}$ and can be satisfied by choosing $s_{00}$ to respect the isometries of FLRW metric. To satisfy the above constraint equation we write down the Lagrangian by following the Arnowitt-Deser-Misner (ADM) formalism proposed in \cite{ONeal-Ault:2020ebv} in which the metric reads
\begin{equation}
ds^2=(\alpha^2-\beta^j\beta_j)dt^2+2\beta_j dtdx^j+\gamma_{ij}dx^idx^j,
\end{equation} 
using which the relevant Lagrangian density is written as   
\begin{equation}\label{eq:Lagrred}
\begin{aligned}
	\mathcal{L} &= \frac{\alpha\sqrt{\gamma}}{2\kappa}\Bigg[\mathcal{R}+(1-\frac{s_{00}}{\alpha^2})\left(K_{ij}K^{ij}-K^2\right) \\&+\frac{2K}{\alpha^2}s_{00}a^ia_i\left(\frac{2}{\alpha^4}s_{00}(\dot{\alpha}-\alpha\beta^ia_i)-\frac{1}{\alpha^3}\dot{s}_{00}\right)\Bigg].
\end{aligned}
\end{equation}
Here,  \( \mathcal{R}\) denotes the three-dimensional scalar Ricci tensor,  \( a_\mu = n^\nu \nabla_\nu n_\mu \) is the ADM acceleration.  The extrinsic curvature \( K_{\mu\nu} \) is defined as  $K_{\mu\nu} = -\nabla_\mu n_\nu - n_\mu a_\nu$ and equivalently expressed via the Lie derivative of the spatial metric \( \gamma_{\mu\nu} \) along the normal vector \( n^\mu \) as $K_{\mu\nu} = -\tfrac{1}{2} \mathcal{L}_n \gamma_{\mu\nu}$. For the presentation of the ADM Lagrangian (\ref{eq:Lagrred}), used of two assumptions. First, just the coefficient \( s_{\mu\nu} \) is non-zero. Second, just one component of the tensor \( s_{\mu\nu} \) is non-zero i.e., $s_{00}$, and the rest ($s_{ij},~~i,j=1,2,3$) are zero.

Imposing the condition that $s_{00}$ remains constant in cosmic time modifies the gravitational dynamics at both background and perturbative levels. Under this constraint along with adding a matter stress-energy tensor of perfect-fluid form $(T_M)^{\mu}_{~\nu}=\text{diag}(-\rho,p,p,p)$ and imposing that it is conserved together with $(T_s)^\mu_{~\nu}$ \cite{ONeal-Ault:2020ebv}, the first Friedmann equation in the FLRW metric (\ref{eq:flrw}) takes the form
\begin{equation}\label{eq:friedt}
	\frac{H^2}{H_0^2} = \Omega_{m0} a^{-3} + \Omega_{r0} a^{-4 x_r} + \Omega_{\Lambda0} a^{x_\Lambda},
\end{equation}
with $x_r = (1-\tfrac{3}{4}s_{00})/(1-\tfrac{1}{2}s_{00})$ and $x_\Lambda=3s_{00}/(1-\tfrac{5}{2}s_{00})$, which arise from the Eq.~(\ref{eq:bianchi}) and lead to a modified continuity equation
\begin{equation}\label{co}
	\dot{\rho}+3\frac{\dot{a}}{a}f(w,s_{00})\rho=0,
\end{equation}
where $f(w,s_{00})=2(1+w-s_{00})/(2+s_{00}(3w-2))$ and where we have neglected the contribution of spatial curvature $\Omega_k^0 a^{-2}$. In these equations, common factors of $s_{00}$ have been absorbed into the definition of $\Omega_X$, and no re-scaling of the time coordinate or the scale factor can fully eliminate its effects. The full set of equations can be found in the Appendix of Ref.~\cite{Nilsson:2022mzq}.
It is important to note that the absence of a pure cosmological constant turns this sector into a slowly evolving type of dynamical dark energy, and was analyzed in~\cite{Nilsson:2023exc}.

\section{Modified strain amplitude for gravitational waves}\label{sec:perts}
The power spectrum of primordial tensor perturbations generated during inflation serves as the key quantity characterizing inflation-driven PGW. In this section, we investigate the impact of explicit spacetime-symmetry breaking on the strain amplitude of the PGW signal. For simplicity, we assume that the tensor perturbations satisfy \( h_{00} = h_{0i} = 0 \). This allows us to work in the transverse traceless (TT) gauge, where the conditions \( \partial^i h_{ij} = 0 \) and \( h^i_i = 0 \) are imposed. Within this framework, the dynamics of the tensor perturbations are described by linear perturbation theory, as outlined in \cite{Nilsson:2022mzq} 
\beq
\label{hdyn}
\ddot h_{ij} + \big(3H-\frac{\dot{s}_{00}}{1-s_{00}}\big) \dot h_{ij}- \frac{\nabla^2}{a^2 (1-s_{00})}h_{ij} = 0
\eeq
This equation reveals two significant modifications compared to the standard GR case. First, the friction (damping) term is modified by the factor $-\dot{s}_{00}/(1-s_{00})$. Second, the effective speed of propagation of tensor modes is altered by the factor $1/(1-s_{00})$. For the case where $s_{00}$ is a constant, the effect of diffeomorphism violation appears just via the third term in Eq. \eqref{hdyn} as well as through the background equations. Also, Eq. (\ref{hdyn}) exhibits a singularity at \(s_{00} = 1\), which corresponds to a strong-coupling limit where the effective kinetic term for tensor modes vanishes. In what follows, we restrict to \(|s_{00}| \ll 1\), consistent with perturbative treatment and observational constraints. For \(s_{00} > 1\), the coefficient \((1 - s_{00})^{-1}\) becomes negative, leading to a gradient instability and exponential growth of tensor modes. Such values are excluded by stability requirements as well as observations.

To solve Eq. \eqref{hdyn}, we transform to Fourier space, following the notation of \cite{Watanabe:2006qe}
\beq
h_{ij}(t,\vec{x}) = \sum_{\lambda}\int\frac{d^3k}{\left(2\pi\right)^3}\,h^\lambda(t,\vec{k})\,\epsilon^\lambda_{ij}(\vec{k})\,e^{i\vec{k}\cdot\vec{x}}\,,
\eeq
where $\epsilon^\lambda_{ij}$ is the spin-2 polarization tensor satisfying the orthonormality condition $\sum_{ij}\epsilon^\lambda_{ij}\epsilon^{\lambda'*}_{ij}=2\delta^{\lambda\lambda'}$, with $\lambda=+,\times$ denoting the two independent polarization states.
The Fourier components $h^\lambda(t,\vec{k})$ can be factorized as
\beq
h^\lambda(t,\vec{k}) = h_{\mathrm{prim}}^\lambda(\vec{k})X(t,k)\,,
\eeq
where $k \equiv |\vec{k}|$ is the wavenumber, $X(t,k)$ represents the transfer function governing the temporal evolution of the perturbation, and $h_{\mathrm{prim}}^\lambda$ characterizes the primordial tensor perturbation amplitude. This parametrization enables us to express the tensor power spectrum as~\cite{Bernal:2019lpc,Bernal:2020ywq}
\beq
\mathcal{P}_T(k) = \frac{k^3}{\pi^2}\sum_\lambda\Big|h^\lambda_{\mathrm{prim}}(\vec k)\Big|^2 = \frac{2}{\pi^2}G \, H^2\Big|_{k=aH}\,.
\eeq
where $G$ is Newton's gravitational constant.
Consequently, the modified evolution equation \eqref{hdyn} for the transfer function in conformal time $d\tau$ (defined via $dt=a d\tau$) becomes
\begin{equation}\label{XGae}
	X'' + 2\mathcal{H}X' + k_{\rm eff}^2X = 0\,,
\end{equation}
where $\mathcal{H}=a H$. This is the equation for a damped harmonic oscillator with a modified, $s_{00}$-dependent effective wavenumber $k_{\rm eff}=k \sqrt{1-s_{00}}$. For a flat universe dominated by a perfect fluid, the scale factor evolves as $a(\tau) \propto \tau^{2/(1+3w)}$, yielding the damping term
\begin{equation}\label{damp}
	2\mathcal{H} = \frac{4}{\tau(1+3w)}\,,
\end{equation}
where $w$ is the equation-of-state parameter, which may receive modifications from \(s_{00}\) as implied by the modified continuity Eq. (\ref{co}). As a result, $s_{00}$ modifies the effective propagation speed of gravitational waves (appearing as 
$k^2 (1-s_{00})$) in the wave equation) but does not affect the Hubble damping term $2\mathcal{H}X'$ in conformal time.

The relic density of PGWs from first-order tensor perturbations in the synchronous gauge is given by \cite{Bernal:2020ywq,Watanabe:2006qe}:
\beq
\Omega_{\text{GW}}(\tau, k) = \frac{[X'(\tau, k)]^2}{12 a^2(\tau) H^2(\tau)} \mathcal{P}_T(k).
\eeq
Assuming rapid oscillation after horizon crossing and averaging over periods, we use the WKB solution: for a mode well inside the horizon, we have \( X(\tau, k) \propto e^{i k_{\text{eff}} \tau} / a(\tau) \) which implies that
\beq
X'(\tau, k) \approx i k_{\text{eff}} X(\tau, k) \quad \Rightarrow \quad |X'(\tau, k)| \approx k_{\text{eff}} |X(\tau, k)|.
\eeq
The horizon crossing (''hc'' subscript) condition is defined when the (modified) mode frequency equals the expansion rate:
\beq
\frac{k_{\text{eff}}}{a(\tau_{\text{hc}})} = H(\tau_{\text{hc}}) \quad \Rightarrow \quad k \sqrt{1 - s_{00}} = a_{\text{hc}} H_{\text{hc}},
\eeq
through which the dependence of $s_{00}$ can be absorbed into the definition of horizon crossing for 
a given physical wavenumber \( k \). Using the WKB amplitude \( |X(\tau, k)| \propto 1/a(\tau) \) and normalizing at horizon crossing, we find:
\beq \label{xa}
|X'(\tau, k)| \approx k_{\text{eff}} |X(\tau, k)| \approx \frac{k_{\text{eff}} a_{\text{hc}}}{\sqrt{2} a(\tau)} = \frac{a_{\text{hc}}^2 H_{\text{hc}}}{\sqrt{2} a(\tau)},
\eeq
where we note that the factor \( \sqrt{1-s_{00}} \) cancels out here because \( k_{\text{eff}} = 2\pi f=a_{\text{hc}} H_{\text{hc}} \). Substituting this into the expression for \( \Omega_{\text{GW}} \) yields
\beq \label{Ttps}
\Omega_{\text{GW}}(\tau, k) \simeq \left[ \frac{a_{\text{hc}}}{a(\tau)} \right]^4 \left[ \frac{H_{\text{hc}}}{H(\tau)} \right]^2 \frac{\mathcal{P}_T(k)}{24},
\eeq
and we find the present-day PGW relic density as
\begin{equation}\label{Spt}
\begin{aligned}
	\Omega_{\mathrm{GW}}(\tau_0,k)h^2 \simeq \left[\frac{g_*(T_{\mathrm{hc}})}{2}\right]&\left[\frac{g_{*s}(T_0)}{g_{*s}(T_{\mathrm{hc}})}\right]^{4/3} \\&\times\frac{\mathcal{P}_T(k)\Omega_{r}(T_0)h^2}{24}\,,
\end{aligned}
\end{equation}
where $h \equiv H_0/(100\,\mathrm{km\,s^{-1}\,Mpc^{-1}})$ is the dimensionless Hubble parameter, $\Omega_r \equiv \rho_r/\rho_{\mathrm{cr}}$ is the radiation density parameter (with $\rho_{\mathrm{cr}} = 3H_0^2/8\pi G$), and $g_*(T)$ and $g_{*s}(T)$ are the effective relativistic degrees of freedom for energy and entropy densities, respectively
\begin{equation}
	\rho_r = \frac{\pi^2}{30}g_*(T)T^4\,, \quad s_r = \frac{2\pi^2}{45}g_{*s}(T)T^3\,.
\end{equation}

The scale dependence of the tensor power spectrum is parameterized as
\begin{equation}\label{pri}
	\mathcal{P}_T(k) = A_T\left(\frac{k}{\tilde k}\right)^{n_T},
\end{equation}
where $n_T$ is the tensor spectral index, $\tilde k = 0.05\,\mathrm{Mpc}^{-1}$ is the Planck 2018 pivot scale. The spectral index $n_T$ characterizes the scale dependence: $n_T = 0$ corresponds to a scale-invariant spectrum, $n_T > 0$ indicates a blue-tilted spectrum, and $n_T < 0$ describes a red-tilted spectrum \cite{Vagnozzi:2023lwo}. The tensor amplitude $A_T$ relates to the scalar amplitude $A_S$ through the tensor-to-scalar ratio
\begin{equation}
	r \equiv \frac{\mathcal{P}_T(\tilde k)}{\mathcal{P}_S(\tilde k)},
\end{equation}
where $\mathcal{P}_S(\tilde k) \simeq 2.21\times10^{-9}$ is the scalar power spectrum amplitude measured by Planck at the CMB scale \cite{Planck:2018vyg}.
The most stringent current constraint on the tensor-to-scalar ratio comes from a joint analysis of Planck CMB data (2018), BICEP/Keck Array observations (BK18), and Baryon Acoustic Oscillation (BAO) data, yielding a 95\% confidence upper limit of \( r_{0.05} < 0.032 \) at the pivot scale \( \tilde{k} = 0.05  \text{Mpc}^{-1} \) \cite{Tristram:2021tvh} \footnote{Note that this bound is further strengthened by its consistency with independent data from the Atacama Cosmology Telescope (ACT), which provides precise measurements of small-scale polarization.}. This is tighter than the previous constraint $r_{0.05}< 0.07$ \cite{BICEP2:2018kqh}.

To examine the effect of the diffeomorphism violation coefficient $s_{00}$ on the PGW spectrum, we can write \(H(\tau) = H_{\mathrm{GR}}(\tau) \cdot \frac{H(\tau)}{H_{\mathrm{GR}}(\tau)}\), resulting in
\beq
\left[ \frac{H_{\mathrm{hc}}}{H(\tau)} \right]^2
= \left[ \frac{H_{\mathrm{hc}}}{H_{\mathrm{GR}}(\tau)} \right]^2 \left[ \frac{H_{\mathrm{GR}}(\tau)}{H(\tau)} \right]^2.
\eeq
Substituting into Eq. \eqref{Ttps}, we have
\beq \label{3}
\Omega_{\mathrm{GW}}(\tau,k) \simeq \left[ \frac{a_{\mathrm{hc}}}{a(\tau)} \right]^4 \left[ \frac{H_{\mathrm{hc}}}{H_{\mathrm{GR}}(\tau)} \right]^2 \left[ \frac{H_{\mathrm{GR}}(\tau)}{H(\tau)} \right]^2 \frac{\mathcal{P}_T(k)}{24},
\eeq
which in general relativity can be re-expressed as follows
\beq \label{2}
\Omega^{\mathrm{GR}}_{\mathrm{GW}}(\tau,k) \simeq \left[ \frac{a^{\mathrm{GR}}_{\mathrm{hc}}}{a^{\mathrm{GR}}(\tau)} \right]^4 \left[ \frac{H^{\mathrm{GR}}_{\mathrm{hc}}}{H^{\mathrm{GR}}(\tau)} \right]^2 \frac{\mathcal{P}_T(k)}{24}.
\eeq
where by inserting $\frac{\mathcal{P}_T(k)}{24}$ from \eqref{2} into Eq. \eqref{3}, we come to
\bea
\Omega_{\mathrm{GW}}(\tau,k) \simeq& \Omega^{\mathrm{GR}}_{\mathrm{GW}}(\tau,k) \left[ \frac{H_{\mathrm{GR}}(\tau)}{H(\tau)} \right]^2 \left[ \frac{a_{\mathrm{hc}}}{a^{\mathrm{GR}}_{\mathrm{hc}}} \right]^4  \left[ \frac{a^{\mathrm{GR}}(\tau)}{a(\tau)} \right]^4 \times \nonumber \\
&\left[ \frac{H_{\mathrm{hc}}}{H^{\mathrm{GR}}_{\mathrm{hc}}} \right]^2,
\eea
where $H_{\mathrm{GR}}(\tau)/H(\tau)$ is the modification from the background expansion at time $\tau$, $a_{\mathrm{hc}}/a^{\mathrm{GR}}_{\mathrm{hc}}$ is the difference in horizon-crossing scale factor, $a^{\mathrm{GR}}(\tau)/a(\tau)$ is the difference in scale factor evolution from horizon crossing to time $\tau$, and $H_{\mathrm{hc}}/H^{\mathrm{GR}}_{\mathrm{hc}}$ is the difference in Hubble parameter at horizon crossing. Also , we use the superscript ''GR`` to denote quantities calculated within general relativity, with $\Omega^{\mathrm{GR}}_{\mathrm{GW}}(\tau, k)$ representing the conventional GR prediction for the PGW relic density (equivalent to Eq. \eqref{Ttps}). At the present epoch ($\tau=0=z$), where $E(z)=1$ as established in
the following definition $H^2=H_0^2 E(z)$
where
\begin{equation}\label{eq:EZ}
E(z)=\Omega_{m0} (1+z)^{3} +\Omega_{r0} (1+z)^{4 x_r}+\Omega_{\Lambda0} (1+z)^{-x_\Lambda},
\end{equation}
and the GW energy density simplifies to

\begin{equation}
	\Omega_{\mathrm{GW}}(\tau_0,k) \simeq \Omega^{\mathrm{GR}}_{\mathrm{GW}}(\tau_0,k)\left[ \frac{a_{\mathrm{hc}}}{a_{\mathrm{hc}}^{\mathrm{GR}}}\right]^4\left[ \frac{H_{\mathrm{hc}}}{H_{\mathrm{hc}}^{\mathrm{GR}}}\right]^2.
	\label{eq:PGWBar} 
\end{equation}
Here, the factors $a_{\mathrm{hc}}/a_{\mathrm{hc}}^{\mathrm{GR}}$, and $H_{\mathrm{hc}}/H_{\mathrm{hc}}^{\mathrm{GR}}$
account for the modified background expansion history due to $s_{00}$, which alters the time of horizon crossing for a given mode wavenumber $k$.

In the end of this section, some comments is essential. Although the tensor perturbation equation (\ref{hdyn}) contains an effective wavenumber \(k_{\mathrm{eff}} = k\sqrt{1-s_{00}}\), the net effect on the present-day energy density \(\Omega_{\mathrm{GW}}\) is dominated by changes in the background expansion history through \(a_{\mathrm{hc}}\) and \(H_{\mathrm{hc}}\), rather than by a modified propagation speed. This is because the factor \(\sqrt{1-s_{00}}\) cancels in the horizon-crossing condition used to derive Eq. (\ref{eq:PGWBar}).
Even though we focus on tensor perturbations and their detectability via GW observatories, the modified background expansion induced by \(s_{00}\) will also affect scalar perturbations and thus CMB observables such as the scalar spectral index \(n_s\), the amplitude of temperature anisotropies, and the locations of acoustic peaks. While a detailed CMB analysis is left for future work, we expect that large values of \(|s_{00}|\) could be constrained by existing CMB data.

\section{Observability of the modified spectrum}\label{sec4}
\begin{figure*}[ht]
%	\begin{center}
		\epsfig{file=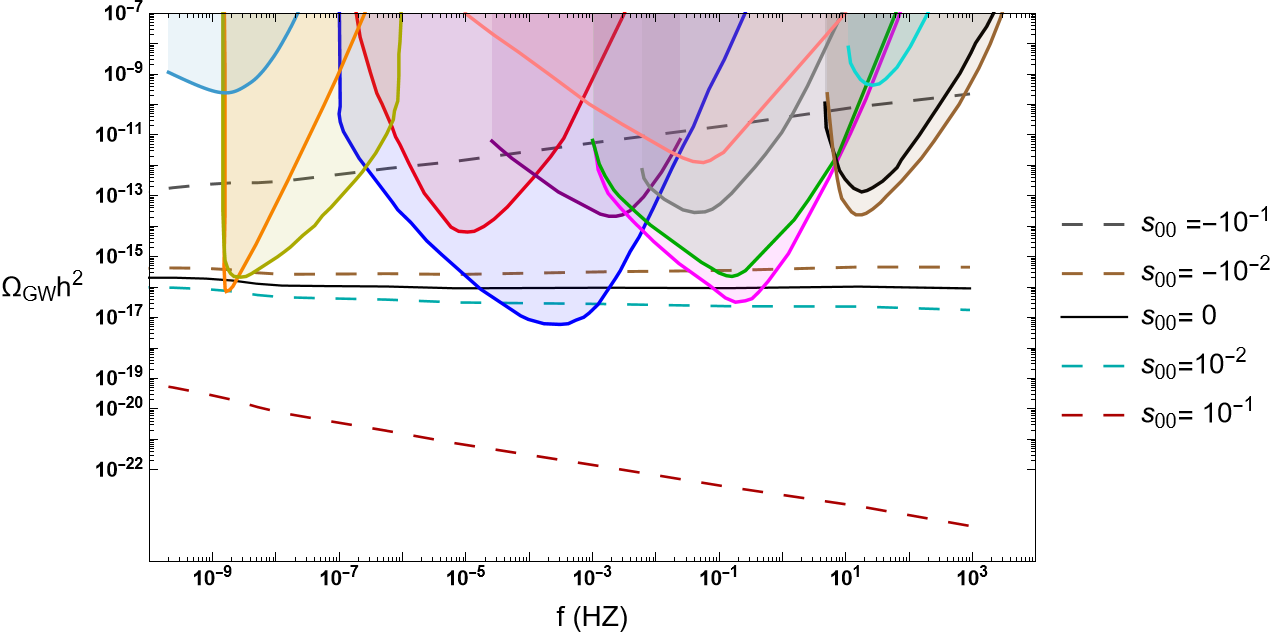,width=0.8\textwidth,angle=0}~~~
		\caption{Plot of the strain amplitude of $\Omega_{GW}h^2$ versus frequency range $10^{-10}~\mathrm{Hz}\leq f\leq 10^4~\mathrm{Hz}$ for the several values of $s_{00}$ within both negative and positive ranges. The colored shaded areas denote different sensitivity regions for upcoming GW detectors: \textbf{NANOGrav} (indigo curve), \textbf{SKA} (orange curve), \textbf{THEIA} (yellow-dark curve), \textbf{$\mu$--Ares} (blue curve), \textbf{ASTROD--GW} (red curve), \textbf{LISA} (purple curve), \textbf{BBO} (magenta curve), \textbf{DECIGO }(green curve), \textbf{CE} (brown curve), \textbf{AION--km} (pink curve), \textbf{AEDGE} (gray curve), \textbf{ET} (black curve), \textbf{aLIGO} (cyan curve).
	 \label{fig:1}}
%	\end{center}
\end{figure*}

In this section, we concentrate on GWs within the frequency range of $(10^{-10}~\mathrm{Hz}, 10^4~\mathrm{Hz})$. Contributions from frequencies below $10^{-10}~\mathrm{Hz}$, which stem from the free streaming of neutrinos and photons~\cite{Weinberg:2003ur}, are excluded. Note that as shown in Ref. \cite{Weinberg:2003ur}, a cosmological background featuring an anisotropic stress tensor, like that of a free-streaming thermal massless neutrino background, can suppress PGWs once they cross the horizon. This finding was further generalized in Ref. \cite{Dent:2013asa} to encompass cases involving massive neutrinos, extra neutrino species, and a possible relativistic background of axions.
The selected frequency range is expected to be comprehensively investigated by both current and future GW observatories, such as NANOGrav \cite{NANOGrav:2023gor}, SKA \cite{Janssen:2014dka}, THEIA \cite{Theia:2019non}, $\mu$-Ares \cite{Sesana:2019vho}, ASTRO-GW \cite{Ni:2012eh}, AION-Km \cite{Badurina:2019hst}, LISA~\cite{LISA:2017pwj}, DECIGO \cite{Kawamura:2020pcg}, AEDGE \cite{AEDGE:2019nxb}, BBO \cite{Crowder:2005nr}, ET \cite{Sathyaprakash:2012jk}, CE \cite{Evans:2021gyd}, and aLIGO \cite{LIGOScientific:2014pky}.
Using Eq. \eqref{eq:PGWBar} and considering the values of \( s_{00} \) in both negative and positive ranges, we present the PGW spectrum, \(\Omega_{\mathrm{GW}}h^2\), as a function of frequency \( f \) (within the previously specified allowed frequency ranges) in Fig. \ref{fig:1}. 
%\onecolumngrid
%
\twocolumngrid
The power spectrum shown in Fig.~\ref{fig:1} is computed using Eq.~\eqref{eq:PGWBar}, which requires establishing the relationship between frequency and redshift through the conversion \(k = H(z)/(1+z)\). By numerically inverting this equation, we obtain \( H_{\text{hc}} \), \( H_{\text{hc}}^{\text{GR}} \), \( a_{\text{hc}} \), and \( a_{\text{hc}}^{\text{GR}} \) as functions of \( k \).  Computationally, we first construct a vector containing pairs of \( z \) and \( k \) values, which allows us to define a numerical function \( k(z) \). Subsequently, we generate another vector combining \( k(z) \) with the ratio \( \Omega_{\text{GW}}(\tau_0, k(z)) / \Omega_{\text{GW}}^{\text{GR}}(\tau_0, k(z)) \), as defined in Eq.~\eqref{eq:PGWBar}. For our numerical computations, we adopt the numerical values $H_0 = 100 \, h_0 = 67.4$ km s$^{-1}$ Mpc$^{-1}$, $\Omega_m^0 = 0.044$, $\Omega_r^0 = 9 \times 10^{-5}$, $T_0 = 2.725$ K. It is worth emphasizing that the fixed values chosen for \( s_{00} \) are intentionally broad to clearly illustrate the impact of \( s_{00} \) on the strain of PGWs.

An important clarification is required regarding the frequency-redshift relation. The connection between the frequency \( f \) and redshift \( z \) is established by the equation \( 2\pi f = a_{\text{hc}} H_{\text{hc}} \), as referenced immediately below Eq.~\eqref{xa}. Since both \( a \) and \( H \) can be expressed as functions of \( z \), this relation allows for a numerical mapping between \( f \) and \( z \), despite the absence of an analytical closed-form solution. 
At first glance, it is evident that when \( s_{00} < 0 \), there is a constructive contribution to the power spectrum of PGWs, whereas no such contribution occurs for \( s_{00} > 0 \) (see Fig.~\ref{fig:1}). 
For positive values of the diffeomorphism-violating parameter \( s_{00} \) exceeding \( 2 \times 10^{-2} \), the modified gravitational wave spectra lie completely below the sensitivity curves of current detectors, making them observationally inaccessible. While in the case where $s_{00} < 0$, this is not the case.
Overall, the PGW spectrum, generated by explicit spacetime symmetry breaking with a parameter \( s_{00} <2\times 10^{-2} \), extends into negative values and intersects the underlying sensitivity curves. 

The primordial tensor perturbations generated during inflation are characterized by the tensor power spectrum \(\mathcal{P}_T(k)\), which at the CMB pivot scale \(\tilde{k} = 0.05\ \text{Mpc}^{-1}\) is parameterized by the amplitude \(A_T\) and the primordial tensor spectral index \(n_T\) as given in Eq.~(\ref{pri}). In the absence of modified background evolution (i.e., within standard \(\Lambda\)CDM), the present-day gravitational-wave energy-density spectrum \(\Omega_{\mathrm{GW}}h^{2}(f)\) can be approximated on large scales by the relation \cite{Chongchitnan:2006pe,Stewart:2007fu}
\begin{equation}
\Omega_{\mathrm{GW}}h^{2}(f)\; \simeq \;4.36\times 10^{-15}\, r\left(\frac{f}{f_{0}}\right)^{\,n_{T}},
\end{equation}
where \(f_{0}=3.1\times 10^{-18}\,\text{Hz}\) is a reference frequency and \(r\equiv A_{T}/A_{S}\) is the tensor-to-scalar ratio. In our model, however, the observed spectrum is not simply of this power-law form; indeed, the explicit diffeomorphism violation encoded in \(s_{00}\) modifies the background expansion history, which in turn alters the transfer function which relates the primordial spectrum to the spectrum today. The full expression for the present-day spectrum is given by Eq.~(\ref{eq:PGWBar}),
where the ratios \(a_{\mathrm{hc}}/a^{\mathrm{GR}}_{\mathrm{hc}}\) and \(H_{\mathrm{hc}}/H^{\mathrm{GR}}_{\mathrm{hc}}\) depend on \(s_{00}\) and introduce an additional scale dependence beyond the primordial tilt \(n_T\). Consequently, the effective slope of \(\Omega_{\mathrm{GW}}h^{2}(f)\) as a function of frequency is a combination of the primordial index \(n_T\) and the transfer-function tilt induced by \(s_{00}\).
Figure~\ref{fig:1} shows the resulting spectra for several values of \(s_{00}\), assuming a fixed primordial spectral index (e.g., \(n_T \approx 0\) or a small red tilt consistent with slow-roll inflation). The plots demonstrate that negative values of \(s_{00}\) enhance the high-frequency part of the spectrum relative to the GR case, mimicking a blue-tilted signal, whereas positive values of \(s_{00}\) suppress it, mimicking a redder tilt. This does not mean that the primordial \(n_T\) itself changes; rather, the observed spectral shape is modified by the background evolution.
\begin{figure*}[ht]
	\begin{tabular}{c}
		%	\centering
		\includegraphics[width=1\columnwidth]{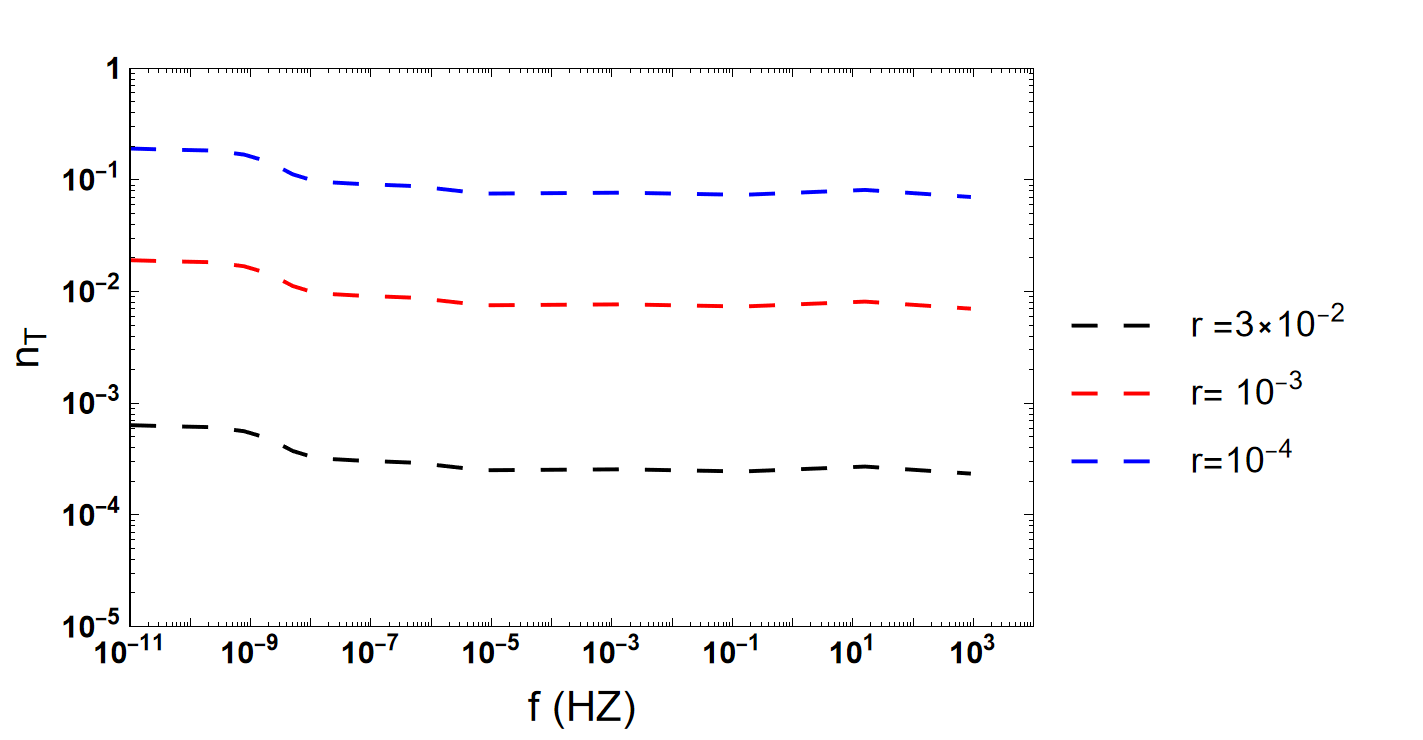}~~~
		\includegraphics[width=1\columnwidth]{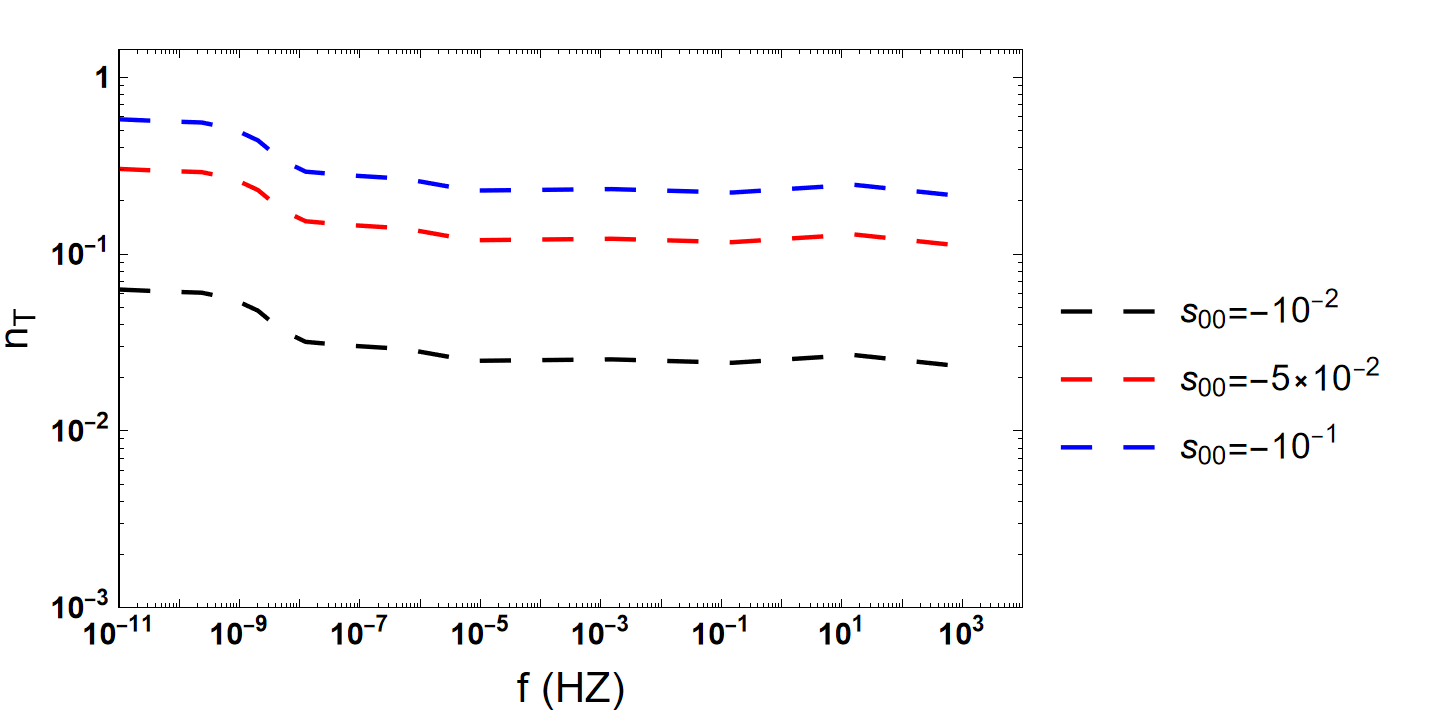}
	\end{tabular}
	\caption{The plot of primordial tensor spectral index $n_T$ in terms of frequency $f$ for some selected values of $r$ with fixed value of $s_{00}=-10^{-4}$ (left) and some selected values of $s_{00}$ with fixed value of $r=0.03$ (right).\label{fig:2}}
	\label{A}
\end{figure*}

\begin{table*}[ht]
	\begin{center}
		{\hfill
			\hbox{
				\begin{tabular}{c c c c}
                \toprule
					$f$ [Hz]&$n_T(r=3\times 10^{-2})$&$n_{T}(r=10^{-2})$&$n_{T}(r=10^{-3})$\\
                    \midrule
					$10^{-11}$&$-0.7$&$-2.12$&$-21$\\
					$5.2\times10^{-11}$&$0.69$&$-2.07$&$-20.7$\\			
					$2.3\times10^{-10}$&$-0.67$&$-2.03$&$-20.3$\\			
					$7.8\times10^{-10}$&$-0.62$&$-1.87$&$-18.7$\\			
					$2\times10^{-9}$&$-0.53$&$-1.62$&$-16.1$\\	
                    \midrule
					$5\times10^{-9}$&$-0.41$&$-1.24$&$-12.4$\\			
					$1.2\times10^{-8}$&$-0.35$&$-1.07$&$-10.7$\\			
					$5.8\times10^{-8}$&$-0.34$&$-1.02$&$-10.2$\\			
					$5.5\times10^{-7}$&$-0.32$&$-0.97$&$-9.77$\\			
					$7.1\times10^{-6}$&$-0.27$&$-0.83$&$-8.3$\\	
                    \midrule
					$10^{-3}$&$-0.28$&$-0.85$&$-8.5$\\			
					$14\times10^{-2}$&$-0.272$&$-0.81$&$-8.1$\\			
					$16$&$-0.3$&$-0.9$&$-9.04$\\			
					$913$&$-0.2$&$-0.78$&$-7.8$\\
                    \bottomrule
				\end{tabular}
			}~
				\hfill
	\hbox{
					\begin{tabular}{c c c c}
                    \toprule
						$f$ [Hz]&$n_T(s_{00}=10^{-1})$&$n_{T}(s_{00}=5\times10^{-2})$&$n_{T}(s_{00}=10^{-2})$\\
						\midrule
						$10^{-11}$&$-1.06$&$-0.5$&$-0.096$\\
						$5.2\times10^{-11}$&$-1.03$&$-0.49$&$-0.094$\\			
						$2.3\times10^{-10}$&$-1.01$&$-0.48$&$-0.092$\\			
						$7.8\times10^{-10}$&$-0.93$&$-0.44$&$-0.085$\\			
						$2\times10^{-9}$&$-0.80$&$-0.38$&$-0.073$\\	
                        \midrule
						$5\times10^{-9}$&$-0.69$&$-0.29$&$-0.056$\\			
						$1.2\times10^{-8}$&$-0.53$&$-0.25$&$-0.048$\\			
						$5.8\times10^{-8}$&$-0.51$&$-0.24$&$-0.046$\\			
						$5.5\times10^{-7}$&$-0.48$&$-0.23$&$-0.044$\\			
						$7.1\times10^{-6}$&$-0.41$&$-0.19$&$-0.038$\\	
                        \midrule
						$10^{-3}$&$-0.42$&$-0.20$&$-0.038$\\			
						$14\times10^{-2}$&$-0.40$&$-0.19$&$-0.037$\\			
						$16$&$-0.45$&$-0.21$&$-0.041$\\			
						$913$&$-0.39$&$-0.18$&$-0.035$\\	
                        \bottomrule
					\end{tabular}
				}
				\hfill
			}
			\caption{Numerical values of primordial $n_{T}$ in terms of $f$ for some selected values of $r$ with fixed value of $s_{00}=10^{-1}$ (left) and some selected values of $s_{00}$ with fixed value of $r=0.02$ (right).}
			\label{Nu}
		\end{center}
	\end{table*}
    
   \begin{table}[h!]
   	\centering
   	\caption{Threshold values of $s_{00}$ for detection by future GW observatories with their maximum sensitivities in terms of $\Omega_{\rm GW}h^2$ and assuming $r = 0.03$ as well as standard cosmological parameters. Thresholds for $s_{00} > 0$ (typically above $2\times10^{-2}$) are not shown.}
   	\label{tab:thresholds}
   	\begin{tabular}{lcc}
   		\toprule
   		Detector & Min[$\Omega_{\rm GW}h^2$] & $s_{00}$ \\
   		\midrule
   		NANOGrav/SKA & $\sim 10^{-9}$ & $\approx -0.06$ \\
   		THEIA & $\sim 10^{-13}$ & $\approx -0.01$ \\
   		$\mu$-Ares & $\sim 10^{-16}$ & $\approx -4 \times 10^{-3}$ \\
   		LISA & $\sim 10^{-12}$ & $\approx -5 \times 10^{-4}$ \\
   		ASTROD-GW & $\sim 10^{-14}$ & $\approx -2 \times 10^{-3}$ \\
   		DECIGO/BBO & $\sim 10^{-15}$ & $\approx -3 \times 10^{-3}$ \\
   		ET/CE & $\sim 10^{-13}$ & $\approx -8 \times 10^{-3}$ \\
   		aLIGO & $\sim 10^{-9}$ & $\lesssim -0.1$ \\
   		\bottomrule
   	\end{tabular}
   \end{table} 

\begin{figure*}[ht]
	\begin{tabular}{c}
		%	\centering
		\includegraphics[width=1\columnwidth]{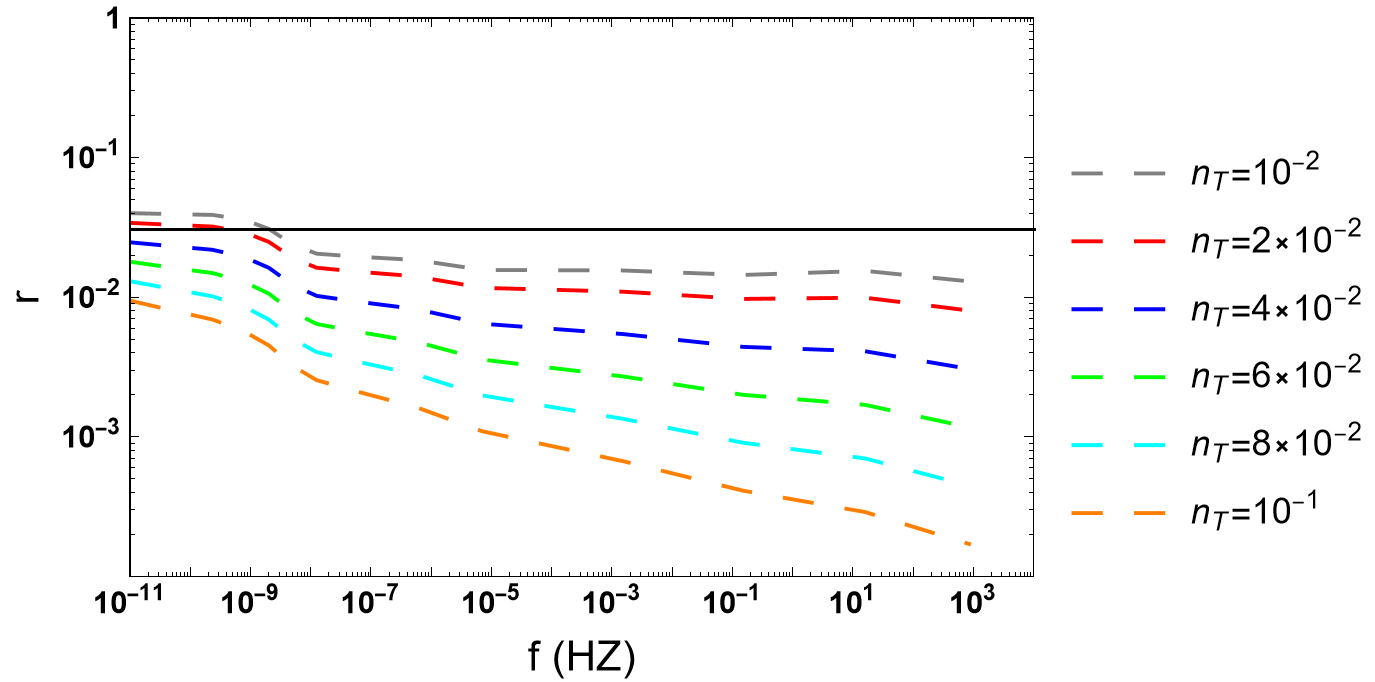}~~~
	\includegraphics[width=1\columnwidth]{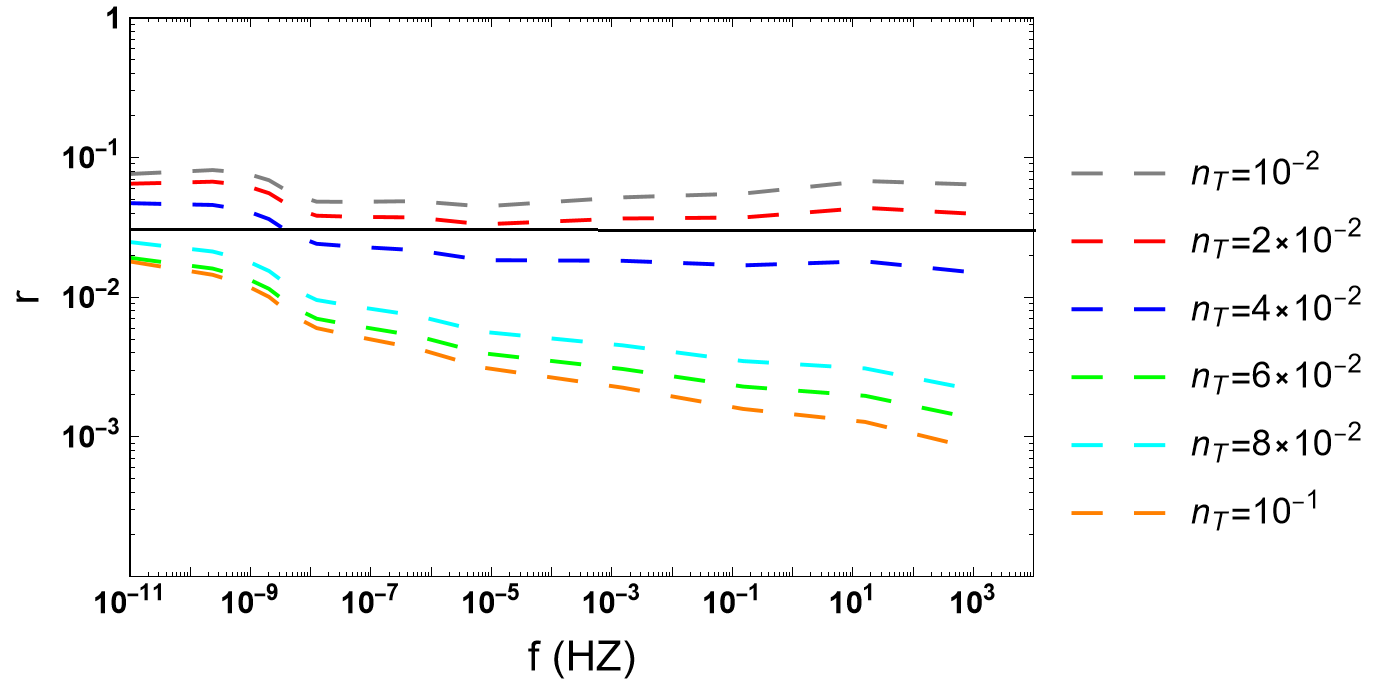}
	\end{tabular}
	\caption{The plot of tensor-to-scalar ratio $r$ in terms of frequency $f$ for some selected values of primordial $n_T$ with fixed values of $s_{00}=-10^{-3}$, and $-10^{-2}$ for left and right panels, respectively. The horizontal black line shows the upper bound $r\lesssim 0.032$. \label{fig:r}}
	\end{figure*}
Given the phenomenological interest in blue-tilted spectra (which are not predicted by standard single-field slow-roll inflation), we illustrate in Fig.~\ref{fig:2} how different choices of the primordial \(n_T\) and \(s_{00}\) affect the observed spectrum. Table~\ref{Nu} summarizes, for selected values of \(s_{00}\), the corresponding range of effective spectral slopes in the frequency bands relevant to various detectors. Figure~\ref{fig:r} then shows, for a fixed primordial \(r\), the range of frequencies at which the signal could be detectable for given values of \(s_{00}\) and \(n_T\). For example, a value \(s_{00} = -10^{-2}\) together with a primordial \(n_T \gtrsim 4\times 10^{-2}\) yields an observed spectrum that rises sufficiently at high frequencies to be accessible by pulsar-timing arrays. Smaller negative values (e.g., \(s_{00} = -10^{-3}\)) require a larger primordial blue tilt (\(n_T \gtrsim 10^{-2}\)) to achieve a comparable enhancement. For \(-10^{-3} < s_{00} < 0\), the effect on the observed spectrum is negligible within the frequency bands shown.
Therefore, the parameter \(s_{00}\) does not alter the primordial tensor index \(n_T\), but it does modify the transfer function, thereby changing the observed frequency dependence of \(\Omega_{\mathrm{GW}}h^{2}(f)\). This transfer-function effect can mimic a blue tilt when \(s_{00}<0\), opening a window for detectability in high-frequency gravitational-wave experiments even if the primordial tensor spectrum is nearly scale-invariant.

\subsection{Threshold values of $s_{00}$ for detectability}
To provide quantitative guidance for future observational searches, we determine the threshold values of the diffeomorphism-violating parameter \( s_{00} \) required for the modified PGW spectrum to cross the sensitivity curves of key upcoming detectors. Determining the exact values of \( s_{00} \) that make the signal detectable by specific instruments turns the qualitative Fig. \ref{fig:1} into a quantitative, testable prediction. This is achieved by solving the equation \( \Omega_{\text{GW}}(f, s_{00}) = \Omega_{\text{detector}}^{\text{sens}}(f) \) at the characteristic frequency where each detector has its peak sensitivity. In order to compute this, we rewrite Eq.~\eqref{eq:PGWBar} as
	\beq
	\Omega_{\text{GW}}(f, s_{00}) = \Omega_{\text{GW}}^{\text{GR}}(f) \times \mathcal{R}(f, s_{00}),
	\eeq
    where \( \mathcal{R}(f, s_{00}) = (a_{\text{hc}}/a_{\text{hc}}^{\text{GR}})^4 (H_{\text{hc}}/H_{\text{hc}}^{\text{GR}})^2 \) is the modification factor, and where \( \Omega_{\text{GW}}^{\text{GR}}(f) \) is calculated using Eq. \eqref{Spt} with a fixed \( r \), which we choose as \( r=0.03 \). For a given detector at its peak frequency \( f_{\text{peak}} \), we solve for \( s_{00} \):
	\beq
	\Omega_{\text{GW}}^{\text{GR}}(f_{\text{peak}}) \times \mathcal{R}(f_{\text{peak}}, s_{00}) = \Omega_{\text{det}}^{\text{sens}}
	\eeq
Since \( \mathcal{R}(f, s_{00}) \) is a monotonic function of \( s_{00} \) (increasing as \( s_{00} \) becomes more negative), this has a unique solution. The results for all detectors are summarized in Table~\ref{tab:thresholds}. 
    
For detectors like NANOGrav and SKA, which probe the nHz frequency band, a relatively large negative value of \( s_{00} \approx -0.06 \) is required to produce a sufficiently blue-tilted spectrum for detection. In the decihertz band, targeted by DECIGO and BBO, the threshold becomes more sensitive, with \( s_{00} \approx -3 \times 10^{-3} \). For the LISA mission in the millihertz band, an even smaller violation of \( s_{00} \approx -5 \times 10^{-4} \) would be sufficient for a detectable signal. Notably, for positive values of \( s_{00} \gtrsim 2 \times 10^{-2} \), the PGW spectrum is suppressed beyond the reach of all planned detectors across the entire frequency range. 
Overall, the detectability thresholds in Table \ref{tab:thresholds} let us compare them with existing constraints on \( s_{00} \). The most stringent bound comes from the multimessenger event GW170817, which constrained the speed of gravitational waves at low redshifts to be nearly identical to the speed of light, translating to \( |s_{00}| \lesssim 10^{-15} \) \cite{LIGOScientific:2017zic}. While this late-time bound is astronomically tighter, it does not necessarily apply to the primordial universe. Our analysis reveals a vast, phenomenologically rich window for large, negative values of \( s_{00} \) (\( -0.08 \lesssim s_{00} \lesssim -10^{-4} \)) in the early Universe that is currently unconstrained and would be accessible to next-generation detectors. A confirmed detection of a stochastic background by, for example, LISA or BBO, that aligns with a blue-tilted spectrum, could be interpreted as a signature of primordial diffeomorphism violation with \( s_{00} \approx -5\times 10^{-4} \) or \( -3\times 10^{-3} \), respectively, orders of magnitude larger than the late-time bound. Conversely, non-detection by these instruments would allow us to place entirely new, early-Universe constraints on \( s_{00} \) at the levels indicated in Table \ref{tab:thresholds}.

These threshold values makes the type of explicit diffeomorphism violation considered here more easily testable: a confirmed detection of a stochastic GW background by any of these instruments that aligns with the \( \Lambda\)CDM prediction would allow us to place an upper bound on \( s_{00} \) at the corresponding level. Conversely, a detection of a blue-tilted spectrum in the nHz band, inconsistent with astrophysical foregrounds, could be interpreted as a signature of a negative \( s_{00} \) of order \( 10^{-2} \).

\section{Big-Bang nucleosynthesis constraints}\label{sec:BBN}
Like all massless degrees of freedom, GWs contribute to the radiation energy density in the early universe. This can have significant implications for key cosmological epochs, such as BBN and recombination, when the CMB was formed. In other words, this contribution affects the expansion rate of the early universe, which in turn influences the freeze-out of nuclear reactions that determine the primordial abundances of light elements (e.g., deuterium and helium-4). The PGWs affect the radiation energy budget, quantified by its contribution to the effective number of relativistic species, \( N_{\rm eff} \). The current value in $\Lambda$CDM is $N_{\rm eff}\approx3.044$ \cite{Froustey:2020mcq,Bennett:2020zkv} but additional GW backgrounds introduce a correction as \( \Delta N_{\text{eff}} \) \cite{Allen:1996vm,Boyle:2007zx} (see also \cite{Vagnozzi:2020gtf,Benetti:2021uea,Giare:2022wxq,Vagnozzi:2022qmc}) through
\begin{eqnarray}
	\Delta N_{\rm eff} \approx 1.8 \times 10^5\int^{f_{\max}}_{f_{\min}}df\,\frac{\Omega_{\rm gw}(f)h^2}{f}\,.
	\label{eq:neff}
\end{eqnarray}
The integration limits, \( f_{\min} \) and \( f_{\max} \), are determined by the specific cosmological epoch under consideration and the maximum temperature achieved during the hot Big Bang era. In this analysis, we focus on the BBN epoch, when light elements were formed. This choice sets \( f_{\min} \simeq 10^{-10} \, \text{Hz} \), corresponding to the frequency of a mode entering the horizon at BBN, when the radiation bath had a temperature of \( T \sim \mathcal{O}(\text{MeV}) \). The contribution to the radiation energy density at any given time comes exclusively from subhorizon modes, as these are the ones that oscillate and behave as propagating massless degrees of freedom. Consequently, frequencies associated with superhorizon modes at the time of interest must be excluded, which is done through the lower limit \( f_{\min} \) in the integral in Eq.~(\ref{eq:neff}). Assuming instantaneous reheating and a standard post-inflationary thermal history, the upper frequency cutoff \( f_{\max} \) is determined by the reheating temperature \( T_{\rm rh} \). For example, a Grand Unified Theory (GUT)-scale reheating temperature corresponds to \( f_{\max} \simeq 10^8 \, \text{Hz} \), with lower \( T_{\rm rh} \) values shifting \( f_{\max} \) downward. To remain consistent with successful BBN, reheating must occur at temperatures \( T_{\rm rh} \gtrsim 5 \, \text{MeV} \)~\cite{deSalas:2015glj}.  

Current BBN and CMB observations constrain the effective extra relativistic degrees of freedom \( \Delta N_{\rm eff} \), with a representative \( 2\sigma \) upper limit of \( \Delta N_{\rm eff} \lesssim 0.4 \)~\cite{Aver:2015iza,Cooke:2017cwo,Planck:2018vyg,Vagnozzi:2019ezj,Hsyu:2020uqb,ACT:2020gnv,Mossa:2020gjc}. However, this work focuses exclusively on BBN constraints for two key reasons. First, perturbation behavior: most CMB analyses model additional radiation as a neutrino-like (free-streaming) fluid~\cite{Vagnozzi:2017ovm,Vagnozzi:2018jhn,RoyChoudhury:2019hls,Giare:2020vzo,Gariazzo:2022ahe}, whereas GW-induced perturbations exhibit distinct dynamics. This raises uncertainties in directly applying CMB-derived \( N_{\rm eff} \) limits to GWs.  Second, spectrum sensitivity: given the strongly blue-tilted spectrum considered here, the integral in Eq.~(\ref{eq:neff}) is effectively insensitive to the lower bound \( f_{\min} = 10^{-10} \, \text{Hz} \) (set by BBN horizon-crossing modes) but highly dependent on \( f_{\max} \). Thus, BBN constraints are more reliable for our purposes of providing conservative estimates.

\begin{figure}[ht]
	\begin{tabular}{c}
		%	\centering
		\includegraphics[width=0.85\columnwidth]{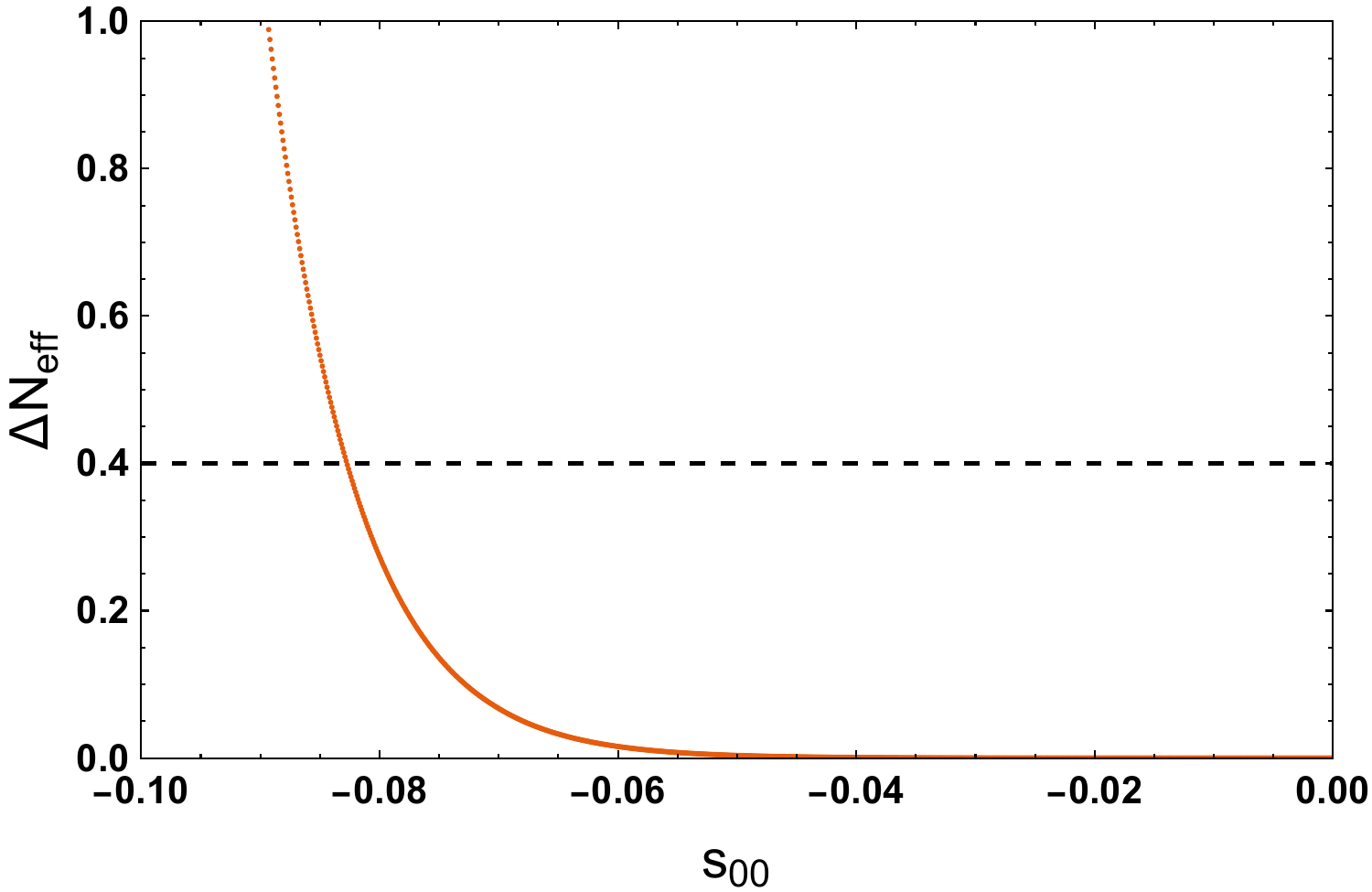}~~~~~~~~~
		\end{tabular}
	\caption{The behavior of $\Delta N_{\rm eff}$ in terms of $s_{00}$. The horizonal dashed line shows the upper bound of $\Delta N_{\rm eff}\lesssim 0.4$. \label{fig:3}}
	\label{AA}
\end{figure}

Diffeomorphism-violating modifications (e.g., nonzero $s_{00}$) may leave imprints in BBN observables via PGW spectra. By substituting the best-fit analytical approximation of the PGW spectrum, \(\Omega_{\text{GW}}h^2(f)\), derived in the previous section into Eq. (\ref{eq:neff}), we numerically evaluate the effective relativistic degrees of freedom, \(\Delta N_{\text{eff}}\), as a function of \(s_{00}\) (see Fig.~\ref{fig:3}).
As shown, the effective relativistic degrees of freedom \(\Delta N_{\text{eff}}\) impose a lower bound on \(s_{00}\), constraining it to \(s_{00} \gtrsim -0.08\). Comparing this constraint with the results in Fig.~\ref{fig:1} reveals that modified PGW spectra within the range $$\boxed{-0.08 \lesssim s_{00} \leq 0}$$ intersect with sensitivity curves of several next-generation detectors as SKA, THEIA, $\mu$-Ares, ASTROD-GW, LISA, DECIGO, and BBO. This intersection represents a phenomenologically significant region that could be probed by future gravitational wave observations.

\section{Discussion \& Conclusions}\label{sec:disc}
In this paper, we have studied the possibility for detecting explicit diffeomorphism violation in primordial tensor modes using current and future gravitational-wave detectors. Starting from a simple model of diffeomorphism violation with only one extra free parameter compared to $\Lambda$CDM, both the background and perturbations and modified, we computed the modified PGW power spectrum as well as the modifications to the tensor-to-scalar ratio and the tensor spectral index, both of which are observables accessible through complementary probes, e.g. CMB experiments. We find that in order for any contribution from diffeomorphism violation to be observable, the coefficient $s_{00}$ needs to be approximately smaller than $\sim0.01$ and can be negative, with aLIGO leading the break at $s_{00}\lesssim-0.1$, with a larger negative value being more observable and a larger positive values being less observable. At the same time, we found that $s_{00}<0$ ($>0$) directly induces a red (blue)-tilted spectrum. Our analysis shows that a value of \( s_{00} \approx -10^{-2} \) can produce a blue tilt of \( n_T \gtrsim 4 \times 10^{-2} \), which could be probed by pulsar timing arrays. Current CMB data, which are consistent with a slightly red-tilted or scale-invariant tensor spectrum and impose a tight upper limit on its amplitude (\( r_{0.05} < 0.032 \)), would be violated by a significant blue tilt. This therefore disfavors large negative values of \( s_{00} \), a conclusion that is consistent with the independent bound we derive from \( \Delta N_{\text{eff}} \).

Many studies on explicit diffeomorphism violation in gravity have been published in the recent years, including in cosmology, where it was for example found that the resulting dynamical dark-energy component breaks the null energy condition and becomes phantom-like (at early times) for $s_{00}>0$~\cite{Nilsson:2023exc}. Moreover, the time evolution of $s_{00}$, which we have turned off in this work, was related to the running of the Planck mass in \cite{Nilsson:2022mzq}, and using projectors, the time-time component of $s^{\mu\nu}$\footnote{Which in the explicit-breaking scenario corresponds to a different theory.} was shown to mimic Milne-like evolution \cite{Reyes:2022dil}. The primordial universe has so far not been the subject of many papers (with the exception of \cite{Nilsson:2022mzq}, of which the present work is a complement). In this paper, a strong constraint on $s_{00}$ was found from the LIGO/Virgo event GW170817 with associated electromagnetic counterpart GRB170817A. From this event, the speed of gravitational waves at late times was found to be $-3\cdot10^{-15}<c_{\rm GW}-1<+7\cdot10^{-16}$ which translates to the bound $-6\cdot10^{-15}<s_{00}<+1.4\cdot10^{-15}$ \cite{LIGOScientific:2017zic}. This is quite a bit stronger than the bounds found in this paper; however, they are not comparable for two reasons. First, the only constraint we have obtained from data is that coming from $\Delta N_{\rm eff}$ and bounds the coefficient as $s_{00}\gtrsim-0.08$, which is consistent with constraints from the speed of gravitational waves. The parameter space for \(s_{00}\) is further restricted by theoretical consistency conditions such as the absence of ghosts and gradient instabilities, which require \(|s_{00}| \ll 1\). This is consistent with our focus on small values of \(s_{00}\) in the observable window \(-0.08 \lesssim s_{00} \lesssim 0.02\).
The other constraints are statements of {\it observability} rather than being direct constraints on $s_{00}$ itself. Second, one may worry that such a strong constraint as that derived from $c_{\rm GW}$ makes the present result redundant; however, the bound on $c_{\rm GW}$ is valid at late times at redshift $z<1$ and for modes deep inside the Hubble horizon. Therefore, there is still a large window available for $s_{00}$ to be large at other scales, but this would be made more clear if we allowed $s_{00}$ to depend on time, as well as considering models beyond this simple case. We leave this for future work.

It is necessary to mention some caveats regarding the main findings. First, it is important to note that the modified early-Universe expansion due to extra relativistic species, decaying particles, or modified-gravity theories can alter the transfer function in ways degenerate with \(s_{00}\). Second, we note that a blue-tilted PGW spectrum at high frequencies can also arise in scenarios where the early universe undergoes a phase dominated by a stiff fluid (\(w > 1/3\)), as discussed in Refs.~\cite{Cui:2017ufi,Figueroa:2019paj}. While such models achieve a similar phenomenological outcome through modified background expansion, the underlying mechanism here is rooted in explicit diffeomorphism violation, which also modifies perturbation dynamics and leaves distinct imprints on the tensor-to-scalar ratio and spectral index which could be distinguished with multi-messenger observations. Therefore, a detection of a blue-tilted stochastic GW background by LISA, DECIGO, or pulsar-timing arrays cannot be uniquely attributed to diffeomorphism violation. Instead, such a detection would indicate either a primordial blue tilt or a modified expansion history—or a combination of both; disentangling these possibilities requires complementary cosmological data.
Conversely, non-detection of a stochastic background by next-generation instruments would place joint constraints on \(s_{00}\) and other parameters governing the early-universe expansion, assuming a given primordial tensor spectrum.

The present work provides a first systematic mapping between the diffeomorphism
violation parameter \(s_{00}\) and the observable PGW spectrum. While the effects are degenerate with other physical mechanisms, the derived thresholds offer a concrete target for future GW searches. Should a stochastic background be detected with a spectrum consistent with a blue tilt, our results indicate the level of diffeomorphism violation that could be responsible, motivating further theoretical work to break the degeneracy—for instance, by studying the simultaneous impact of \(s_{00}\) on scalar perturbations and CMB observables, or by allowing time evolution of \(s_{00}\) (which we have set to zero here).

In summary, explicit diffeomorphism violation represents one of several viable pathways to an enhanced high-frequency GW background. Future multi-messenger and multi-frequency cosmological observations will be essential to distinguish among these possibilities and to probe fundamental spacetime symmetries in the early Universe.

\begin{acknowledgments}
N.A.N was supported by the Institute for Basic Science under the project code IBS-R018-D3 and acknowledges support from PSL/Observatoire de Paris. Moreover, the authors networking support from the COST Actions CA21136 - ``Addressing observational tensions in cosmology with systematics and fundamental physics (CosmoVerse)'', supported by COST - ``European Cooperation in Science and Technology'' and CA23130
``Bridging high and low energies in search of quantum gravity (BridgeQG)''.
\end{acknowledgments}

\bibliography{reference}

\end{document}